\documentclass[fleqn,usenatbib]{mnras}
\usepackage{newtxtext,newtxmath}

\usepackage[T1]{fontenc}

\DeclareRobustCommand{\VAN}[3]{#2}
\let\VANthebibliography\thebibliography
\def\thebibliography{\DeclareRobustCommand{\VAN}[3]{##3}\VANthebibliography}

\usepackage{graphicx}
\usepackage{amsmath}

\title[On the isotropy of viscosity in discs]{On the isotropy of viscosity in accretion discs}

\author[Nixon \& Pringle]{
C. J. Nixon,$^{1}$\thanks{E-mail: c.j.nixon@leeds.ac.uk}
and J. E. Pringle$^{2}$
\\
$^{1}$School of Physics and Astronomy, Sir William Henry Bragg Building, Woodhouse Ln., University of Leeds, Leeds LS2 9JT, UK\\
$^{2}$Institute of Astronomy, University of Cambridge, Madingley Road, Cambridge CB3 0HA, UK, and Emmanuel College, Cambridge,  CB2 3AP, UK
}

\date{Accepted XXX. Received YYY; in original form ZZZ}

\pubyear{\the\year{}}

\begin{document}
\label{firstpage}
\pagerange{\pageref{firstpage}--\pageref{lastpage}}
\maketitle

\begin{abstract}
Accretion discs are fundamental to many astrophysical systems, providing the conversion of gravitational potential energy into radiation that we can observe. In many systems there is evidence that discs are warped; from spatially-resolved observations of protoplanetary discs, to the features of lightcurves and line profiles from discs around supermassive black holes in galaxy centres. The dynamics of warped discs is largely controlled by the physical nature of the internal disc viscosity. While typically disc viscosity is hydromagnetic in origin, simulations of magnetized discs cannot match observed rates of angular momentum transport in planar discs and thus cannot be used to determine the ratio of the torques responsible for driving accretion to those responsible for evolving the disc warp. The analytic work of Ogilvie is the most comprehensive model for warped disc evolution, but makes assumptions that need to be tested. In particular, it assumes that the disc viscosity is Navier-Stokes, and therefore small-scale and isotropic. Here we attempt to test this model using the long periods of X-ray binaries that are due to precession of the disc. These systems have well-constrained estimates of the component of viscosity responsible for driving accretion, and by looking at systems with and without evidence for disc misalignment and precession we can constrain the component of viscosity responsible for flattening the disc. We conclude that the observational constraints suggest that the Ogilvie model provides an adequate description of the disc evolution, but that there are indications that the internal disc viscosity might be marginally non-isotropic.
\end{abstract}


\section{Introduction}
\label{intro}
Accretion discs are a fundamental building block in many areas of astronomy. Discs form on a range of scales from around planets to around supermassive black holes in galaxy centres \citep{Pringle:1981}. Accretion discs are the vehicle by which material is transported on to the central object, governing the assembly and growth of stars and generating the spectacular luminosities from distant compact objects. The fundamental mechanism that causes a disc to evolve is typically referred to as a ``viscosity'', as this process is responsible for the exchange of angular momentum between disc orbits and ultimately the energy release and driving of the accretion flow \citep{Pringle:1972,Shakura:1973}. The internal workings of accretions discs, i.e. the physical nature of the viscosity, has been an active topic of current research for several decades and remains so today.

The majority of the attention on disc viscosity has naturally been on the internal stresses responsible for driving accretion in planar discs; i.e. those discs in which all of the disc orbits rotate in the same plane. In this case, and for a geometrically thin disc, we may take the disc orbits to lie in the $z=0$ plane (in cylindrical polar coordinates $R, \phi, z$) rotating with angular velocity $\Omega = \sqrt{GM/R^3}$ where $M$ is the mass of the central object. In this case the only shearing motion is due to $d\Omega/dR$ and is in the $R\phi$ direction. This, through viscosity, leads to an $R\phi$ stress and results in the radially outward transport of angular momentum and the inward transport of mass. The result is a diffusion equation for the evolution of the surface density $\Sigma(R, t)$, and the rate of diffusion is controlled by the kinematic viscosity, $\nu$.

Solving for the subsequent disc evolution requires a model for $\nu$. While it introduces several assumptions that may be more or less satisfied in certain discs, \cite{Shakura:1973} introduced a simple parametrization of $\nu$ in terms of local variables. The model takes the viscosity to be $\nu = \alpha c_{\rm s} H$, where $c_{\rm s}$ is the sound speed, $H$ is the disc semi-thickness, and the dimensionless viscosity parameter $\alpha$ defines the strength of the viscosity. This has been successful in describing the behaviour of many observed systems with a single free parameter. The value that $\alpha$ takes in real discs, and indeed the applicability of a local model, depends on the nature of the viscosity. \cite{Shakura:1973} suggest that on basic physical grounds we expect $\alpha \lesssim 1$.

For gas pressure supported discs that are highly ionized, the value of $\alpha$ is now well-determined from observational considerations. Modeling of the time-dependent behaviour of such discs in Dwarf Novae and Soft X-ray transients leads to the conclusion that $\alpha \approx 0.2 - 0.3$ \citep[][see also \citealt{King:2007,Kotko:2012}]{Martin:2019}. \cite{Martin:2019} also advance arguments in favour of the suggestion by \cite{Shakura:1973} that the viscous process in this case is predominantly hydro-magnetic. In contrast, \cite{King:2007} and \cite{Nixon:2024} draw attention to the problem that numerical simulations of hydro-magnetic turbulence in accretion discs give rise to values of $\alpha$ that are far too small to be consistent with the observations.

So far we have discussed the case of a planar disc in which there is one component to the internal disc shearing motions. However, in many astrophysical systems, discs are expected to be warped to some degree, with the plane of rotation varying with distance from the central object. Study of the dynamics of warped discs has a long history \citep[][see \citealt{Nixon:2016} for a review]{Papaloizou:1983,Papaloizou:1995a,Papaloizou:1995b,Ogilvie:1999}. The evidence for disc warps from observations ranges from indirect evidence from unresolved discs (e.g. the connection of various features in a lightcurve to the dynamics of a warped disc; e.g. \citealt{Pasham:2024}) to the now available direct evidence of spatially resolved observations of nearby protoplanetary discs \citep[e.g.][]{Rosenfeld:2012,Jensen:2014,Casassus:2015,Walsh:2017,Kraus:2020,Villenave:2024}.

The structure of warped discs is determined by the combination of the external torques that are applied to the disc (e.g. by a misaligned companion star) and by the way in which the warp propagates through the disc. The latter is determined by the local properties of the disc, and principally by the disc angular thickness, $H/R$, and the disc viscosity parameter, $\alpha$ \citep{Papaloizou:1983}. In particular, these properties alter the internal stresses generated within a warped disc. While the external torques acting on the disc can be readily calculated from known dynamics, the internal stresses in accretion discs are more elusive. This is, in part, due to the complexity of the problem (there are limited symmetries of the problem that can be exploited; cf. \citealt{Ogilvie:1999}) and in part due to the inherent magnetic nature of the stresses (which makes directly applicable numerical simulation difficult; \citealt{Nixon:2024}). 

The most complete theory of warped discs is provided by \citet[][see also \citealt{Ogilvie:2013}]{Ogilvie:1999,Ogilvie:2000}. This theory extends previous analyses by e.g. \cite{Papaloizou:1983} and \cite{Pringle:1992} to provide a detailed description of the internal torques in a non-magnetic fluid disc, generated by a warp in the presence of a Navier-Stokes (i.e. isotropic) viscosity whose magnitude is assumed to be given by the parameter $\alpha$. This assumption is not naturally expected from magnetic stresses. It is therefore important to try to assess the degree to which the results which stem from the assumption of isotropy agree with the observations. To do this, we extend the analysis of \cite{Nixon:2015} who argued that the assumption of the presence of radiation warping in the system Her X-1 can be used to set a limit to the (an)isotropy. To put this in context, we first, in Section~\ref{warppropagation}, outline the theory of viscous propagation of warps in accretion discs. In Section~\ref{radiationwarping}, we discuss the concept of radiation warping of accretion discs and the limit this can put on the viscous anisotropy. In Section~\ref{eta} we show how consideration of binary systems either side of the radiation warping boundary can be used to provide limits to the viscous anisotropy. We present a summary and conclusions in Section~\ref{conclusions}.

\section{Viscous Warp Propagation}
\label{warppropagation}
How a warp propagates through a disc depends on the internal structure of the disc through the relative size of the disc angular semi-thickness and the rate at which radial excursions of the gas are damped. The vertical stratification of the disc, with the pressure a decreasing function of distance from the midplane, leads to an imbalance in the radial direction where the disc is warped; as the local plane of rotation changes with radius the midplanes are no longer aligned. There is also an imbalance in a planar disc as the pressure at the midplane varies with radius, but this imbalance is azimuthally symmetric and thus the effect is to (weakly) alter the azimuthal velocity at which the disc rotates. In a warped disc, the pressure imbalance takes on an $m=1$ structure, varying around a ring of disc material proportional to $\cos\phi$. This has the effect of introducing epicyclic motion to the disc material, which in a Keplerian disc also takes on an $m=1$ (eccentric) motion. These motions result in additional shearing motions, and thus give rise to viscous stresses, in the $\phi z$ and $Rz$ directions. To compute the magnitude of these stresses it is necessary to make assumptions about the isotropy of the internal disc viscosity. As we have mentioned above, the usual, and simplest,  assumption is that the internal disc viscosity is isotropic.

We here restrict our attention to the high-viscosity discs in which the warp propagation is diffusive, and thus principally a local process.\footnote{Note that here we refer to locality in the sense that the warp propagation is not wavelike, in which warp waves carry information from one location to another. As opposed to the possibility, discussed above, that the internal stresses in the disc are non-local, as would be effected by e.g. magnetic fields structures with radial scales $\sim R \gg H$.} In these discs, for which $\alpha \gg H/R$, the warp is transmitted radially by local viscous stresses which result from the internal dynamics combining the hydromagnetic turbulence with the orbital response of the disc to the presence of the warp.

As in the case of a planar disc, conservation laws can again be used to generate an equation that determines the evolution of the warp in the disc \citep{Pringle:1992}. In the case of a warped disc it is necessary to introduce a second component of the viscosity $\nu_2$ that is responsible for communicating the warp radially through the disc. This viscosity is typically discussed as relating to the $Rz$-stress, but arises through a combination of stresses to generate the communication through the disc of the component of disc angular momentum parallel to the local orbital plane \citep{Papaloizou:1983,Nixon:2015}. In the case that the warping of the disc is not too severe, and  that the second viscosity is significantly larger than the usual viscosity (i.e. $\nu_2 \gg \nu_1$), then the time-dependence of the shape of the warp follows a diffusion equation with the rate of diffusion given by $\nu_2$, and thus the relevant smoothing timescale of a warped disc that is not externally forced is $t_{\rm smooth} \approx R^2/\nu_2$. 

In the case that the internal viscosity is locally isotropic with magnitude given by the parameter $\alpha$, the response of the disc is such that $\nu_1 \ne \nu_2$. \cite{Papaloizou:1983} showed that for small $\alpha$ and small warp amplitudes, $\psi$, the result is $\nu_1 \simeq \alpha c_{\rm s}H$ and $\nu_2 \simeq \tfrac{1}{2\alpha}c_{\rm s}H$ and the ratio is $\eta = \nu_2/\nu_1 = 1/(2\alpha^2)$. \cite{Ogilvie:1999,Ogilvie:2000} provides a more complete analysis, accounting for the nonlinear fluid dynamics, that is valid for arbitrary $\alpha$ and warp amplitudes, which for small warps yields
\begin{equation}
	\eta = \frac{\nu_2}{\nu_1} = \frac{1}{2 \alpha^2} \frac{4(1 + 7 \alpha^2)}{4 + \alpha^2}\,.
\end{equation}
We note that when considering the stability boundary for warping of the disc, it is the viscosity values at small warp amplitudes that is relevant to the analysis.

In Section~\ref{intro}, we noted that the $\alpha$ model of \cite{Shakura:1973} has achieved significant success in describing the behaviour of many disc systems, suggesting that a local model of disc viscosity is at least a good first order approach and that the disc viscosity parameter is of order $\alpha \approx 0.3$. It is therefore plausible that the local model for $\nu_2$ provides a good description of the evolution of warped discs. However, what is less clear is whether the assumption of isotropy is a good one. Previous investigations have addressed this question from different approaches \citep{Torkelsson:2000,Ogilvie:2003,King:2013,Nixon:2015}. Here, we extend the analysis in \cite{Nixon:2015} to additional sources that allow us to constrain the second component of warped disc viscosity. In particular we make use of observational constraints to estimate the magnitude of the quantity $\eta = \nu_2/\nu_1$ (Equation 1), thus providing a test of the assumption that the internal stress-strain relations within the disc can be adequately modelled using an isotropic (Navier-Stokes) viscosity.

\section{Radiation Warping and Application to X-ray binaries}
\label{radiationwarping}
\cite{Pringle:1996,Pringle:1997}, following \cite{Iping:1990}, demonstrated that accretion discs may be warped by radiation from the central object.  Whether or not such a warp occurs depends on whether the radiation torque (which increases the warp) is able to overcome the viscous torque (proportional to $\nu_2$ which tends to reduce the warp). Pringle found that for a steady disc with a constant accretion rate, $\dot{M}$, the disc is susceptible to radiation warping at radii $R > R_{\rm crit}$, where
\begin{equation}
\label{Rcrit}
\frac{R_{\rm crit}}{R_s} = \left( \frac{2}{\gamma_{\rm crit}^2} \right) \frac{\eta^2}{\epsilon^2}.
\end{equation}
Here $R_s = 2GM/c^2$, where $M$ is the mass of the object at the centre of the disc and $\epsilon = L/{\dot M} c^2$, where $L$ is the luminosity of the central object, assumed to be isotropic, and $\eta = \nu_2/\nu_1$. \cite{Pringle:1996} made a simple analytic estimate and suggested $\gamma_{\rm crit} = 1/2\pi$. In \cite{Pringle:1997} he made a more accurate numerical estimate and found that $\gamma_{\rm crit} \approx 0.32$. \cite{Ogilvie:2001} carried out similar numerical computations to \cite{Pringle:1997}. They too (see \citealt{Ogilvie:2001}, Figure 1) found that the original analytic estimate of $\gamma_{\rm crit}$ made by \cite{Pringle:1996} was too large and found values similar to that found by \cite{Pringle:1997}.

\cite{Wijers:1999} applied these ideas to the discs in X-ray binaries. The idea here is that once the disc in such systems becomes susceptible to being warped out of the plane of the binary orbit, the disc itself begins to precess. The precession is driven by a combination of tidal torques from the binary companion and radiation torques acting on the warped disc. Such precession manifests itself observationally as a long-term, or superorbital, period, $P_{\rm long}$,  in the luminosity from the system. For typical X-ray binaries the tidal precession is retrograde, and  is such that $P_{\rm prec} \approx 20 P_{\rm orb}$, where $P_{\rm orb}$ is the binary period (see, for example, \citealt{Bate:2000}, and references therein), The radiation torques give rise to similar  values for $P_{\rm prec}$ but the induced precession can be either prograde or retrograde depending on the disc configuration \citep{Wijers:1999}. 

\cite{Wijers:1999} showed that for a variety of X-ray binaries the rate of precession and the disc tilt obtained for realistic values of system parameters compared favourably with the known body of data on X-ray binaries with long periods. 

\cite{Ogilvie:2001} carried out numerical simulations using the specific model for disc viscosity given in \cite{Ogilvie:2000}. In line with observational findings, they assumed that $\alpha = 0.3$, and thus using the \cite{Ogilvie:2000} model they assumed that $\eta = 8.86$. They found (see \citealt{Ogilvie:2001}, Figure~7) that only the low-mass X-ray binaries with the longest orbital periods ($P_{\rm orb} \ge 1$d) are likely to be unstable to radiation-driven warping. 

\citet[][Figure 1]{Kotze:2012} give a more recent assessment of the presence of super-orbital periods in X-ray binaries and, using the analysis of \cite{Ogilvie:2001} come to similar conclusions. 

\section{Using observations to estimate the value of $\eta = \nu_2/\nu_1$}
\label{eta}

If the value of $\eta$ is known, then from Equation~\ref{Rcrit}, it is evident that discs smaller than a certain value cannot be unstable to radiation warping. As the disc size is determined by the size of the binary orbit, this is equivalent to arguing that X-ray binaries with orbital periods smaller than a certain value cannot be unstable to radiation warping. This is the conclusion reached by \cite{Ogilvie:2001} and by \cite{Kotze:2012}, based on the assumption that the isotropic model adopted by \cite{Ogilvie:2000} for the disc viscosity is correct. 

However, it is perhaps preferable to try to establish the value of $\eta$ directly from the observational data, without using any preconceptions about the viscosity model. It is well established observationally that the value of $\alpha$ for the kind of strongly ionized accretion discs found in X-ray binaries is in the range $\alpha = 0.2 - 0.3$ \citep{Martin:2019}. For the \cite{Ogilvie:2000} viscosity model this would imply $\eta = 15.8 - 8.86$. 

\cite{Sood:2007} investigate super-orbital periods in X-ray binaries, and give in their Table 1 a list of the properties of X-ray binary systems with confirmed super-orbital modulations. In their Figure 1 \cite{Sood:2007} plot the superorbital periods against the orbital periods for these systems. They note that there is a loose correlation between the two for all but the shortest orbital periods. The correlation agrees approximately with the relation $P_{\rm long} \approx 20P_{\rm orb}$ expected from the simulations. It is evident from Equation~\ref{Rcrit} that the most stringent limits on the value of $\eta$ are likely to come from the systems subject to radiation warping that have the shortest orbital periods (to a first approximation, all other things being equal, it is evident from equation~\ref{Rcrit} that the deduced value of $\eta \propto P_{\rm orb}^{1/3}$). 

For this reason, in order to set observational limits to the value of $\eta$, we consider the three short period systems in the list of objects compiled by \cite{Sood:2007} that appear to span the zone which defines the stability/instability boundary to radiation warping. These objects are: (i) UW CrB, the object with the shortest orbital period $(P_{\rm orb} = 0.077\,{\rm d})$ that shows strong evidence for radiation warping; (ii) V1405 Aql with a slightly shorter orbital period $(P_{\rm orb} = 0.035\,{\rm d})$, which shows behaviour that might be interpreted as evidence for intermittent radiation warping; and (iii) 4U 1820-303 which has an orbital period $P = 0.008\,{\rm d}$ and shows no evidence for radiation warping. We discuss each of these in turn.

\subsection{UW Coronae Borealis }
The system in Figure 1 of \cite{Sood:2007} which lies on the $P_{\rm long} \approx 20 P_{\rm orb}$ relationship and has the shortest orbital period is UW CrB (also known as MS1603+2600). \cite{Hakala:2005} report an orbital period of  $P_{\rm orb} = 111\,{\rm min} = 0.077$\,d, and note the presence of Type 1 X-ray bursts indicating that the primary star is a neutron star. \cite{Hakala:2009} report a long period $P_{\rm long} \approx 5$\,d and interpret this as disc precession. \cite{Mason:2012} report on long term observations of UW CrB. They note that the new observations are consistent with a model in which the accretion disc in UW CrB is asymmetric and precesses in the prograde direction with a period of $\approx 5.5$\,days. They also suggest that one can estimate the mass of the secondary star by assuming that it fits the mass-radius relation for secondary stars in cataclysmic variables \citep{Patterson:2005} which would give $M_2 = 0.14 M_\odot$. If we assume that the mass of the neutron star is the Chandrasekhar limit, $M_1 = 1.4 M_\odot$, this would give a mass ratio of $q = 0.1$. Using these numbers we find that the binary separation is $a = 6.13 \times  10^{10}$\,cm, a volume average Roche lobe radius of $R_{\rm L} = 0.58 a$ \citep{Eggleton:1983}, and hence a tidal truncation radius for the disc of $R_{\rm t} = 0.87 R_{\rm L}$ \citep{Papaloizou:1977}. 

Using these parameters, and if we also assume that the long period in this case is indeed due to radiation warping of the disc, we then require that the disc radius be less than the critical radius, i.e. that  $R_{\rm t} < R_{\rm crit}$. Using Equation~\ref{Rcrit} with $\gamma_{\rm crit} = 0.32$ and $\epsilon = 0.1$ we now find that we require $\eta \le 6.15$. We note that this is close to, but marginally outside, the range suggested by the \cite{Ogilvie:2000} model. 

\subsection{V1405 Aquilae}

The system XB1916-053 (V1405 Aql) has an orbital period of $P_{\rm orb} = 0.035$\,d, which is a factor of $\sim 2$ less than that of UW CrB. A long period of $P_{\rm long} = 5$\,d was reported by \cite{Homer:2001} and noted by \cite{Charles:2008}, along with an even longer modulation period of 199\,d. \cite{Kotze:2012} confirmed the presence of the 199 d modulation but found no evidence for the 5\,d period. \cite{Sood:2007} plot only the 199\,d period in their Figure 1, and note that this period lies well outside the loose correlation of $P_{\rm long} \approx 20 P_{\rm orb}$. Thus, at best, this system shows only intermittently a period which might be interpreted as due to radiation warping.  In this case, it might be that the intermittency of the 5\,d period in this system (if confirmed) indicates that in this system $R_{\rm t}  \approx R_{\rm crit}$. If so, using the same analysis as above, but taking account of the smaller orbital period would suggest that   $\eta \approx 5$.

\subsection{4U 1820-303}

The system 4U 1820-303 has an orbital period of $P_{\rm orb} = 0.008$\,d and shows a long term modulation period of $P_{\rm long} \approx 172 - 176$\,d \citep{Sood:2007,Wang:2010,Kotze:2012}. This long term period has been interpreted as being caused by the third body in the system \citep{Chou:2001}. In any case, it seems clear that radiation warping is not present in this system. \cite{Wang:2010} suggest that the system consists of a neutron star with mass $M_1 = 1.4 M_\odot$ receiving mass from a white dwarf of mass $M_2 = 0.06 - 0.08 M_\odot$. Using the same analysis as before, but now requiring that the critical radius $R_{\rm crit}$ lie outside the radius of the disc, $R_{\rm t}  >R_{\rm crit}$, we find that this would imply $\eta > 3.65$.

\bigskip

\section{Summary and Conclusions}
\label{conclusions}

Using simple conservation equations \citep[][see also \citealt{Papaloizou:1983}]{Pringle:1992} derived evolution equations for a warped disc. He required two parameters, with dimensions of kinematic viscosity: $\nu_1$ (or more usually simply $\nu$) which describes the timescale for redistribution of disc surface density, and $\nu_2$ which describes the evolution of the disc warp. \citet[][see also \citealt{Papaloizou:1983}]{Ogilvie:2000} using the equations of fluid dynamics, incorporating a Navier-Stokes viscosity $\nu$, showed that the equations proposed by \citep{Pringle:1992} were essentially correct and derived relationships between the quantities $\nu_2$ and $\nu$. Of particular note is that typically we expect $\nu_2 \gg \nu$, so that the timescale for the damping of a warp is expected to be much shorter than the timescale for the flow of mass through the disc. For small warps this relationship is given in Equation~1.  In this paper we attempt to use observational results to assess the validity of this relationship.

To do this, we make use of the finding \citep{Pringle:1996,Pringle:1997} that in binary stars which involve flow through a disc onto a compact object, it is possible that feedback from the accretion luminosity can warp the disc. The warped disc also precesses, leading to an observed ``superorbital period'' some twenty times longer than the orbital period. For this to be able to occur the disc needs to be large enough (Equation~2), and as a corollary it cannot occur in systems that are too small or, equivalently, have orbital periods that are too short.

In this paper we have examined those systems listed by \cite{Sood:2007} with the shortest orbital periods, which display superorbital periods. We have in mind that for such systems, if the superorbital period is due to radiation warping then we expect $P_{\rm long} \approx 20 P_{\rm orb}$. We know that for the discs in such systems the usual viscosity $\nu$ corresponds to $\alpha \approx 0.2 - 0.3$ \citep{Martin:2019}. Using Equation~1, which is based on the assumption that that the viscosity is Navier-Stokes (i.e. isotropic), we might therefore expect $\eta$ to be in the range $\eta \approx 15.8 - 8.9$. We find three such systems which are likely close to the stability limit. One shows a superorbital period in the right range, one appears to do so sporadically, and one does not. Using the known system properties to model the disc sizes, and then using Equation 2, we find that $\eta$ is likely in the range $\eta \approx 3.65 -  6.15$ with a best guess of $\eta \approx 5$. 

Taken at face value, our main finding is that we confirm the expectation that $\nu_2 \gg \nu$. We find, however, that the value of $\nu_2$ is marginally smaller than is predicted by disc models which assume a Navier-Stokes (isotropic) viscosity. If so, this is in the direction suggested by \cite{Pringle:1992} who pointed out that if the underlying stresses are magnetic then there might be a distinction between the effects of  $R\phi$ shear, which is secular, and $Rz$ shear, which is oscillatory, in the sense that the oscillatory shear might be less dissipative.

We must stress, however, that in this analysis we have  glossed over a host of complications. We have taken  the accretion efficiency throughout to be $\epsilon = 0.1$. Since the accretors here are all neutron stars, taking the same value for them all is not unreasonable. In the modelling of the radiation instability criterion \cite{Pringle:1997}, \cite{Wijers:1999} and \cite{Ogilvie:2001} all make specific assumptions about where in the disc matter is added. For example \cite{Ogilvie:2001} assume matter is added at the circularisation radius, which is unlikely if one is wanting to look at the stability border for warping. All the authors use the assumption that the disc is infinitessimally thin and that the luminosity source is isotropic and emitted from a point of negligible size. In reality these assumptions, while plausible as a first approximation, may give rise to enough uncertainty that the apparent discrepancies between the expected and observed values of $\eta$ should not be taken as definitive.

\section*{Acknowledgements}
CJN acknowledges support from the Leverhulme Trust (grant no. RPG-2021-380).

\section*{Data Availability}
The data underlying this article can be made available upon reasonable request to the author.

\bibliographystyle{mnras}
\bibliography{nixon} 

\begin{thebibliography}{}
\makeatletter
\relax
\def\mn@urlcharsother{\let\do\@makeother \do\$\do\&\do\#\do\^\do\_\do\%\do\~}
\def\mn@doi{\begingroup\mn@urlcharsother \@ifnextchar [ {\mn@doi@}
  {\mn@doi@[]}}
\def\mn@doi@[#1]#2{\def\@tempa{#1}\ifx\@tempa\@empty \href
  {http://dx.doi.org/#2} {doi:#2}\else \href {http://dx.doi.org/#2} {#1}\fi
  \endgroup}
\def\mn@eprint#1#2{\mn@eprint@#1:#2::\@nil}
\def\mn@eprint@arXiv#1{\href {http://arxiv.org/abs/#1} {{\tt arXiv:#1}}}
\def\mn@eprint@dblp#1{\href {http://dblp.uni-trier.de/rec/bibtex/#1.xml}
  {dblp:#1}}
\def\mn@eprint@#1:#2:#3:#4\@nil{\def\@tempa {#1}\def\@tempb {#2}\def\@tempc
  {#3}\ifx \@tempc \@empty \let \@tempc \@tempb \let \@tempb \@tempa \fi \ifx
  \@tempb \@empty \def\@tempb {arXiv}\fi \@ifundefined
  {mn@eprint@\@tempb}{\@tempb:\@tempc}{\expandafter \expandafter \csname
  mn@eprint@\@tempb\endcsname \expandafter{\@tempc}}}

\bibitem[\protect\citeauthoryear{{Bate}, {Bonnell}, {Clarke}, {Lubow},
  {Ogilvie}, {Pringle}  \& {Tout}}{{Bate} et~al.}{2000}]{Bate:2000}
{Bate} M.~R.,  {Bonnell} I.~A.,  {Clarke} C.~J.,  {Lubow} S.~H.,  {Ogilvie}
  G.~I.,  {Pringle} J.~E.,   {Tout} C.~A.,  2000, \mn@doi [\mnras]
  {10.1046/j.1365-8711.2000.03648.x}, \href
  {https://ui.adsabs.harvard.edu/abs/2000MNRAS.317..773B} {317, 773}

\bibitem[\protect\citeauthoryear{{Casassus} et~al.,}{{Casassus}
  et~al.}{2015}]{Casassus:2015}
{Casassus} S.,  et~al., 2015, \mn@doi [\apj] {10.1088/0004-637X/811/2/92},
  \href {https://ui.adsabs.harvard.edu/abs/2015ApJ...811...92C} {811, 92}

\bibitem[\protect\citeauthoryear{{Charles}, {Clarkson}, {Cornelisse}  \&
  {Shih}}{{Charles} et~al.}{2008}]{Charles:2008}
{Charles} P.,  {Clarkson} W.,  {Cornelisse} R.,   {Shih} C.,  2008, \mn@doi
  [\nar] {10.1016/j.newar.2008.03.025}, \href
  {https://ui.adsabs.harvard.edu/abs/2008NewAR..51..768C} {51, 768}

\bibitem[\protect\citeauthoryear{{Chou} \& {Grindlay}}{{Chou} \&
  {Grindlay}}{2001}]{Chou:2001}
{Chou} Y.,  {Grindlay} J.~E.,  2001, \mn@doi [\apj] {10.1086/324038}, \href
  {https://ui.adsabs.harvard.edu/abs/2001ApJ...563..934C} {563, 934}

\bibitem[\protect\citeauthoryear{{Eggleton}}{{Eggleton}}{1983}]{Eggleton:1983}
{Eggleton} P.~P.,  1983, \mn@doi [\apj] {10.1086/160960}, \href
  {https://ui.adsabs.harvard.edu/abs/1983ApJ...268..368E} {268, 368}

\bibitem[\protect\citeauthoryear{{Hakala}, {Ramsay}, {Muhli}, {Charles},
  {Hannikainen}, {Mukai}  \& {Vilhu}}{{Hakala} et~al.}{2005}]{Hakala:2005}
{Hakala} P.,  {Ramsay} G.,  {Muhli} P.,  {Charles} P.,  {Hannikainen} D.,
  {Mukai} K.,   {Vilhu} O.,  2005, \mn@doi [\mnras]
  {10.1111/j.1365-2966.2004.08543.x}, \href
  {https://ui.adsabs.harvard.edu/abs/2005MNRAS.356.1133H} {356, 1133}

\bibitem[\protect\citeauthoryear{{Hakala}, {Hjalmarsdotter}, {Hannikainen}  \&
  {Muhli}}{{Hakala} et~al.}{2009}]{Hakala:2009}
{Hakala} P.,  {Hjalmarsdotter} L.,  {Hannikainen} D.~C.,   {Muhli} P.,  2009,
  \mn@doi [\mnras] {10.1111/j.1365-2966.2008.14374.x}, \href
  {https://ui.adsabs.harvard.edu/abs/2009MNRAS.394..892H} {394, 892}

\bibitem[\protect\citeauthoryear{{Homer}, {Charles}, {Hakala}, {Muhli}, {Shih},
  {Smale}  \& {Ramsay}}{{Homer} et~al.}{2001}]{Homer:2001}
{Homer} L.,  {Charles} P.~A.,  {Hakala} P.,  {Muhli} P.,  {Shih} I.-C.,
  {Smale} A.~P.,   {Ramsay} G.,  2001, \mn@doi [\mnras]
  {10.1046/j.1365-8711.2001.04170.x}, \href
  {https://ui.adsabs.harvard.edu/abs/2001MNRAS.322..827H} {322, 827}

\bibitem[\protect\citeauthoryear{{Iping} \& {Petterson}}{{Iping} \&
  {Petterson}}{1990}]{Iping:1990}
{Iping} R.~C.,  {Petterson} J.~A.,  1990, \aap, \href
  {https://ui.adsabs.harvard.edu/abs/1990A&A...239..221I} {239, 221}

\bibitem[\protect\citeauthoryear{{Jensen} \& {Akeson}}{{Jensen} \&
  {Akeson}}{2014}]{Jensen:2014}
{Jensen} E. L.~N.,  {Akeson} R.,  2014, \mn@doi [\nat] {10.1038/nature13521},
  \href {https://ui.adsabs.harvard.edu/abs/2014Natur.511..567J} {511, 567}

\bibitem[\protect\citeauthoryear{{King}, {Pringle}  \& {Livio}}{{King}
  et~al.}{2007}]{King:2007}
{King} A.~R.,  {Pringle} J.~E.,   {Livio} M.,  2007, \mn@doi [\mnras]
  {10.1111/j.1365-2966.2007.11556.x}, \href
  {https://ui.adsabs.harvard.edu/abs/2007MNRAS.376.1740K} {376, 1740}

\bibitem[\protect\citeauthoryear{{King}, {Livio}, {Lubow}  \& {Pringle}}{{King}
  et~al.}{2013}]{King:2013}
{King} A.~R.,  {Livio} M.,  {Lubow} S.~H.,   {Pringle} J.~E.,  2013, \mn@doi
  [\mnras] {10.1093/mnras/stt364}, \href
  {https://ui.adsabs.harvard.edu/abs/2013MNRAS.431.2655K} {431, 2655}

\bibitem[\protect\citeauthoryear{{Kotko} \& {Lasota}}{{Kotko} \&
  {Lasota}}{2012}]{Kotko:2012}
{Kotko} I.,  {Lasota} J.~P.,  2012, \mn@doi [\aap]
  {10.1051/0004-6361/201219618}, \href
  {https://ui.adsabs.harvard.edu/abs/2012A&A...545A.115K} {545, A115}

\bibitem[\protect\citeauthoryear{{Kotze} \& {Charles}}{{Kotze} \&
  {Charles}}{2012}]{Kotze:2012}
{Kotze} M.~M.,  {Charles} P.~A.,  2012, \mn@doi [\mnras]
  {10.1111/j.1365-2966.2011.20146.x}, \href
  {https://ui.adsabs.harvard.edu/abs/2012MNRAS.420.1575K} {420, 1575}

\bibitem[\protect\citeauthoryear{{Kraus} et~al.,}{{Kraus}
  et~al.}{2020}]{Kraus:2020}
{Kraus} S.,  et~al., 2020, \mn@doi [Science] {10.1126/science.aba4633}, \href
  {https://ui.adsabs.harvard.edu/abs/2020Sci...369.1233K} {369, 1233}

\bibitem[\protect\citeauthoryear{{Martin}, {Nixon}, {Pringle}  \&
  {Livio}}{{Martin} et~al.}{2019}]{Martin:2019}
{Martin} R.~G.,  {Nixon} C.~J.,  {Pringle} J.~E.,   {Livio} M.,  2019, \mn@doi
  [\na] {10.1016/j.newast.2019.01.001}, \href
  {https://ui.adsabs.harvard.edu/abs/2019NewA...70....7M} {70, 7}

\bibitem[\protect\citeauthoryear{{Mason}, {Robinson}, {Bayless}  \&
  {Hakala}}{{Mason} et~al.}{2012}]{Mason:2012}
{Mason} P.~A.,  {Robinson} E.~L.,  {Bayless} A.~J.,   {Hakala} P.~J.,  2012,
  \mn@doi [\aj] {10.1088/0004-6256/144/4/108}, \href
  {https://ui.adsabs.harvard.edu/abs/2012AJ....144..108M} {144, 108}

\bibitem[\protect\citeauthoryear{{Nixon}}{{Nixon}}{2015}]{Nixon:2015}
{Nixon} C.,  2015, \mn@doi [\mnras] {10.1093/mnras/stv796}, \href
  {https://ui.adsabs.harvard.edu/abs/2015MNRAS.450.2459N} {450, 2459}

\bibitem[\protect\citeauthoryear{{Nixon} \& {King}}{{Nixon} \&
  {King}}{2016}]{Nixon:2016}
{Nixon} C.,  {King} A.,  2016, in {Haardt} F.,  {Gorini} V.,  {Moschella} U.,
  {Treves} A.,   {Colpi} M.,  eds, , Vol.~905, Lecture Notes in Physics.
Berlin Springer Verlag, p.~45, \mn@doi{10.1007/978-3-319-19416-5_2}

\bibitem[\protect\citeauthoryear{{Nixon}, {Pringle}  \& {Pringle}}{{Nixon}
  et~al.}{2024}]{Nixon:2024}
{Nixon} C.~J.,  {Pringle} C.~C.~T.,   {Pringle} J.~E.,  2024, \mn@doi [Journal
  of Plasma Physics] {10.1017/S002237782300140X}, \href
  {https://ui.adsabs.harvard.edu/abs/2024JPlPh..90a9001N} {90, 905900101}

\bibitem[\protect\citeauthoryear{{Ogilvie}}{{Ogilvie}}{1999}]{Ogilvie:1999}
{Ogilvie} G.~I.,  1999, \mn@doi [\mnras] {10.1046/j.1365-8711.1999.02340.x},
  \href {https://ui.adsabs.harvard.edu/abs/1999MNRAS.304..557O} {304, 557}

\bibitem[\protect\citeauthoryear{{Ogilvie}}{{Ogilvie}}{2000}]{Ogilvie:2000}
{Ogilvie} G.~I.,  2000, \mn@doi [\mnras] {10.1046/j.1365-8711.2000.03654.x},
  \href {https://ui.adsabs.harvard.edu/abs/2000MNRAS.317..607O} {317, 607}

\bibitem[\protect\citeauthoryear{{Ogilvie}}{{Ogilvie}}{2003}]{Ogilvie:2003}
{Ogilvie} G.~I.,  2003, \mn@doi [\mnras] {10.1046/j.1365-8711.2003.06359.x},
  \href {https://ui.adsabs.harvard.edu/abs/2003MNRAS.340..969O} {340, 969}

\bibitem[\protect\citeauthoryear{{Ogilvie} \& {Dubus}}{{Ogilvie} \&
  {Dubus}}{2001}]{Ogilvie:2001}
{Ogilvie} G.~I.,  {Dubus} G.,  2001, \mn@doi [\mnras]
  {10.1046/j.1365-8711.2001.04011.x}, \href
  {https://ui.adsabs.harvard.edu/abs/2001MNRAS.320..485O} {320, 485}

\bibitem[\protect\citeauthoryear{{Ogilvie} \& {Latter}}{{Ogilvie} \&
  {Latter}}{2013}]{Ogilvie:2013}
{Ogilvie} G.~I.,  {Latter} H.~N.,  2013, \mn@doi [\mnras]
  {10.1093/mnras/stt916}, \href
  {https://ui.adsabs.harvard.edu/abs/2013MNRAS.433.2403O} {433, 2403}

\bibitem[\protect\citeauthoryear{{Papaloizou} \& {Lin}}{{Papaloizou} \&
  {Lin}}{1995}]{Papaloizou:1995a}
{Papaloizou} J.~C.~B.,  {Lin} D.~N.~C.,  1995, \mn@doi [\apj] {10.1086/175127},
  \href {https://ui.adsabs.harvard.edu/abs/1995ApJ...438..841P} {438, 841}

\bibitem[\protect\citeauthoryear{{Papaloizou} \& {Pringle}}{{Papaloizou} \&
  {Pringle}}{1977}]{Papaloizou:1977}
{Papaloizou} J.,  {Pringle} J.~E.,  1977, \mn@doi [\mnras]
  {10.1093/mnras/181.3.441}, \href
  {https://ui.adsabs.harvard.edu/abs/1977MNRAS.181..441P} {181, 441}

\bibitem[\protect\citeauthoryear{{Papaloizou} \& {Pringle}}{{Papaloizou} \&
  {Pringle}}{1983}]{Papaloizou:1983}
{Papaloizou} J.~C.~B.,  {Pringle} J.~E.,  1983, \mn@doi [\mnras]
  {10.1093/mnras/202.4.1181}, \href
  {https://ui.adsabs.harvard.edu/abs/1983MNRAS.202.1181P} {202, 1181}

\bibitem[\protect\citeauthoryear{{Papaloizou} \& {Terquem}}{{Papaloizou} \&
  {Terquem}}{1995}]{Papaloizou:1995b}
{Papaloizou} J. C.~B.,  {Terquem} C.,  1995, \mn@doi [\mnras]
  {10.1093/mnras/274.4.987}, \href
  {https://ui.adsabs.harvard.edu/abs/1995MNRAS.274..987P} {274, 987}

\bibitem[\protect\citeauthoryear{{Pasham} et~al.,}{{Pasham}
  et~al.}{2024}]{Pasham:2024}
{Pasham} D.~R.,  et~al., 2024, \mn@doi [\nat] {10.1038/s41586-024-07433-w},
  \href {https://ui.adsabs.harvard.edu/abs/2024Natur.630..325P} {630, 325}

\bibitem[\protect\citeauthoryear{{Patterson} et~al.,}{{Patterson}
  et~al.}{2005}]{Patterson:2005}
{Patterson} J.,  et~al., 2005, \mn@doi [\pasp] {10.1086/447771}, \href
  {https://ui.adsabs.harvard.edu/abs/2005PASP..117.1204P} {117, 1204}

\bibitem[\protect\citeauthoryear{{Pringle}}{{Pringle}}{1981}]{Pringle:1981}
{Pringle} J.~E.,  1981, \mn@doi [\araa] {10.1146/annurev.aa.19.090181.001033},
  \href {https://ui.adsabs.harvard.edu/abs/1981ARA&A..19..137P} {19, 137}

\bibitem[\protect\citeauthoryear{{Pringle}}{{Pringle}}{1992}]{Pringle:1992}
{Pringle} J.~E.,  1992, \mn@doi [\mnras] {10.1093/mnras/258.4.811}, \href
  {https://ui.adsabs.harvard.edu/abs/1992MNRAS.258..811P} {258, 811}

\bibitem[\protect\citeauthoryear{{Pringle}}{{Pringle}}{1996}]{Pringle:1996}
{Pringle} J.~E.,  1996, \mn@doi [\mnras] {10.1093/mnras/281.1.357}, \href
  {https://ui.adsabs.harvard.edu/abs/1996MNRAS.281..357P} {281, 357}

\bibitem[\protect\citeauthoryear{{Pringle}}{{Pringle}}{1997}]{Pringle:1997}
{Pringle} J.~E.,  1997, \mn@doi [\mnras] {10.1093/mnras/292.1.136}, \href
  {https://ui.adsabs.harvard.edu/abs/1997MNRAS.292..136P} {292, 136}

\bibitem[\protect\citeauthoryear{{Pringle} \& {Rees}}{{Pringle} \&
  {Rees}}{1972}]{Pringle:1972}
{Pringle} J.~E.,  {Rees} M.~J.,  1972, \aap, \href
  {https://ui.adsabs.harvard.edu/abs/1972A&A....21....1P} {21, 1}

\bibitem[\protect\citeauthoryear{{Rosenfeld} et~al.,}{{Rosenfeld}
  et~al.}{2012}]{Rosenfeld:2012}
{Rosenfeld} K.~A.,  et~al., 2012, \mn@doi [\apj] {10.1088/0004-637X/757/2/129},
  \href {https://ui.adsabs.harvard.edu/abs/2012ApJ...757..129R} {757, 129}

\bibitem[\protect\citeauthoryear{{Shakura} \& {Sunyaev}}{{Shakura} \&
  {Sunyaev}}{1973}]{Shakura:1973}
{Shakura} N.~I.,  {Sunyaev} R.~A.,  1973, \aap, \href
  {https://ui.adsabs.harvard.edu/abs/1973A&A....24..337S} {24, 337}

\bibitem[\protect\citeauthoryear{{Sood}, {Farrell}, {O'Neill}  \&
  {Dieters}}{{Sood} et~al.}{2007}]{Sood:2007}
{Sood} R.,  {Farrell} S.,  {O'Neill} P.,   {Dieters} S.,  2007, \mn@doi
  [Advances in Space Research] {10.1016/j.asr.2007.02.057}, \href
  {https://ui.adsabs.harvard.edu/abs/2007AdSpR..40.1528S} {40, 1528}

\bibitem[\protect\citeauthoryear{{Torkelsson}, {Ogilvie}, {Brandenburg},
  {Pringle}, {Nordlund}  \& {Stein}}{{Torkelsson}
  et~al.}{2000}]{Torkelsson:2000}
{Torkelsson} U.,  {Ogilvie} G.~I.,  {Brandenburg} A.,  {Pringle} J.~E.,
  {Nordlund} {\r{A}}.,   {Stein} R.~F.,  2000, \mn@doi [\mnras]
  {10.1046/j.1365-8711.2000.03647.x}, \href
  {https://ui.adsabs.harvard.edu/abs/2000MNRAS.318...47T} {318, 47}

\bibitem[\protect\citeauthoryear{{Villenave} et~al.,}{{Villenave}
  et~al.}{2024}]{Villenave:2024}
{Villenave} M.,  et~al., 2024, \mn@doi [\apj] {10.3847/1538-4357/ad0c4b}, \href
  {https://ui.adsabs.harvard.edu/abs/2024ApJ...961...95V} {961, 95}

\bibitem[\protect\citeauthoryear{{Walsh}, {Daley}, {Facchini}  \&
  {Juh{\'a}sz}}{{Walsh} et~al.}{2017}]{Walsh:2017}
{Walsh} C.,  {Daley} C.,  {Facchini} S.,   {Juh{\'a}sz} A.,  2017, \mn@doi
  [\aap] {10.1051/0004-6361/201731334}, \href
  {https://ui.adsabs.harvard.edu/abs/2017A&A...607A.114W} {607, A114}

\bibitem[\protect\citeauthoryear{{Wang} \& {Chakrabarty}}{{Wang} \&
  {Chakrabarty}}{2010}]{Wang:2010}
{Wang} Z.,  {Chakrabarty} D.,  2010, \mn@doi [\apj]
  {10.1088/0004-637X/712/1/653}, \href
  {https://ui.adsabs.harvard.edu/abs/2010ApJ...712..653W} {712, 653}

\bibitem[\protect\citeauthoryear{{Wijers} \& {Pringle}}{{Wijers} \&
  {Pringle}}{1999}]{Wijers:1999}
{Wijers} R. A.~M.~J.,  {Pringle} J.~E.,  1999, \mn@doi [\mnras]
  {10.1046/j.1365-8711.1999.02720.x}, \href
  {https://ui.adsabs.harvard.edu/abs/1999MNRAS.308..207W} {308, 207}

\makeatother
\end{thebibliography}

\bsp
\label{lastpage}
\end{document}